\documentclass[useAMS,usenatbib]{mn2e}
\usepackage{graphicx}
\usepackage{bm}

\begin{document}
\title[Made-to-measure galaxy models III]{Made-to-measure galaxy models - III Modelling with Milky Way observations}
\author[R. J. Long, Shude Mao, Juntai Shen and Yougang Wang]
  {R. J.~Long$^{1,2}$\thanks{E-mail: rjl2007@gmail.com}, Shude~Mao$^{1,2}$,  Juntai~Shen$^3$ and Yougang~Wang$^{4,1}$ \\
   $^1$National Astronomical Observatories, Chinese Academy of Sciences, A20 Datun Rd, Chaoyang District, Beijing 100012, China\\
   $^2$Jodrell Bank Centre for Astrophysics, Alan Turing Building, The University of Manchester, Manchester M13 9PL, UK\\
   $^3$Key Laboratory for Research in Galaxies and Cosmology, Shanghai Astronomical Observatory, Chinese Academy of Sciences, \\ 80 Nandan Road, Shanghai 200030, China\\
   $^4$Key Laboratory of Optical Astronomy, National Astronomical Observatories, Chinese Academy of Sciences, Beijing 100012, China}
\date{Accepted 2012 October 24.  Received 2012 October 19; in original form 2012 August 28}
\pagerange{\pageref{firstpage}--\pageref{lastpage}} \pubyear{2012}
\maketitle 

\label{firstpage}

\begin{abstract}
We demonstrate how the Syer \& Tremaine made-to-measure method of stellar dynamical modelling can be adapted to model a rotating galactic bar.  We validate our made-to-measure changes using observations constructed from the existing Shen et al. (2010) N-body model of the Milky Way bar, together with kinematic observations of the Milky Way bulge and bar taken by the Bulge Radial Velocity Assay (BRAVA).  Our results for a combined determination of the bar angle and bar pattern speed ($\approx 30^{\circ}$ and $\approx 40 \: \rm{km/s/kpc}$) are consistent with those determined by the N-body model.  Whilst the made-to-measure techniques we have developed are applied using a particular N-body model and observational data set, they are in fact general and could be applied to other Milky Way modelling scenarios utilising different N-body models and data sets.  Additionally, we use the exercise as a vehicle for illustrating how N-body and made-to-measure methods might be combined into a more effective method.
\end{abstract}


\section{Introduction}\label{sec:introduction}
The made-to-measure method (M2M) proposed by \citet{Syer1996} for modelling stellar dynamical systems has been the subject of growing interest recently (for example, \citealt{EJ2007}, \citealt{DL2007}, \citealt{DL2008a}, \citealt{Dehnen2009}, \citealt{Long2010}, \citealt{Das2011}, \citealt{Morganti2012}).  This paper is the third in a series together with \citet{Long2010} and \citet{Long2012}.  Whereas the previous two papers, in common with other authors, dealt with theoretical galaxy models and external galaxies, this paper applies the M2M method to the Milky Way.  \citet{Bissantz2004} previously used the method to model the equatorial plane surface brightness of the Galaxy.  Note however that the effectiveness of their model is subject to debate \citep{Rattenbury2007}. We believe that the current investigation is the first in which Galactic M2M modelling has taken place using a rotating frame, and with luminosity and kinematic observables based on fields specified in terms of Galactic longitude and latitude.

\citet{Shen2010} investigated whether the Milky Way bulge might be formed from disk and bar related mechanisms.  (\citealt{Shen2010} provide a much fuller analysis of the issues.)  Starting from a simple N-body disk simulation of some $10^6$ self-gravitating particles, a bar develops naturally (from the conventional bar instability), buckles and thickens, and a pseudo-bulge appears.  With appropriate scaling, they are able to match kinematic data taken from the BRAVA (Bulge Radial Velocity Assay) survey \citep{Rich2007, Howard2008, Kunder2012}, and thus suggest that the formation of the Milky Way bulge is mainly shaped by the internal dynamical buckling instability.  Their model weakly constrains the Galactic bar to an angle of $\approx 20^{\circ}$ to the Sun-Galactic centre line, with a pattern speed (J. Shen private communication) of $\approx 40 \: \rm{km/s/kpc}$.  In our M2M modelling, we utilise the \citet{Shen2010} particle data and the BRAVA field observations, and investigate whether the values for the bar angle and pattern speed we determine are consistent with \citet{Shen2010}.  Note that our intention is not to refine or improve upon the N-body results.

The M2M method has been demonstrated to be a flexible tool for modelling with a variety of observables with  measurements distributed in various different spatial configurations.  The method can handle luminous and kinematic observables and there is no requirement that the observable measurements should share a common spatial positioning.  However in order to function it does require a system of particles with a set of initial conditions and a gravitational potential to generate their orbits.  N-body models have been used extensively to model galactic componentry, for example disks, spirals, bars, black holes (\citealt{Hernquist1993}, \citealt{Sigurdsson1995}, \citealt{Sellwood2012}, \citealt{Sellwood1993}, \citealt{Fux1999}, \citealt{Shen2010}).  The models are effective in modelling the general properties of generic galaxies but have limited tailoring capabilities for modelling the kinematics of real galaxies.  Thus, there are potentially advantages to be gained by combining the two modelling methods with the N-body method providing an initial model including particles and potential, and the M2M method refining the model by weighting the particles to match the measured observables.  One might imagine a sequence of interleaved N-body and M2M models with feedback from one model type being used to improve modelling with the other.  By checking consistency or otherwise of results between the two approaches, the current paper is an initial step in this direction.  Note that we will not be utilising interleaving as described earlier but using an existing published N-body model and seeking to use the M2M method to corroborate the N-body model and results.  If corroboration can not be achieved for this simple case then we must understand (and address) the underlying reasons before developing a more tightly coupled approach between the two methods.

To summarise, our objectives in performing this investigation are three-fold:
\begin{enumerate}
\item to update our M2M implementation, and the theory underlying the M2M method if necessary, to support modelling of a rotating galactic bar and to handle Milky Way observations,
\item to validate our M2M changes using the \citet{Shen2010} Milky Way N-body bar model and the BRAVA observations with the expectation that we should achieve results consistent with the N-body model, and
\item to aid our understanding of the synergy that might be developed between coupled N-body and M2M modelling.  
\end{enumerate}
The structure of the paper is as follows. In section \ref{sec:m2m} we describe the M2M method, and in section \ref{sec:milkyway} we consider the changes required to construct and run our M2M bar models. In sections \ref{sec:results} and \ref{sec:conclusions}, we discuss our results and draw conclusions within the context of the objectives.  In section \ref{sec:discussion} we discuss further combining N-body and M2M modelling.

\section{The M2M Method}\label{sec:m2m}

\subsection{Outline}
In brief, the M2M method is concerned with modelling stellar systems and individual galaxies as a system of test particles orbiting in a gravitational potential.  Weights are associated with the particles and are adjusted as the particles are orbited so that, by using these weights, observational measurements of a real galaxy are reproduced.  As conceived and demonstrated by \citet{Syer1996} and extended by \citet{Long2010} (quantifying weight convergence), we expect that the weights themselves will have converged individually to some constant value.  It is natural to relate the particle weights to the luminosity of a galaxy and then to consider how the galaxy's surface brightness and luminosity weighted kinematics could be generated using the particle system.

In the next section, based heavily on \citet{Long2012}, we set out the theory underlying the M2M method.

\subsection{Theory}\label{sec:theory}
For a system of $N$ particles, orbiting in a gravitational potential, with weights $w_i$, the key equation which leads to the weight evolution equation is
\begin{equation}
	F(\bmath{w}) = -\frac{1}{2} \chi ^2 + \mu S + \frac{1}{\epsilon}\frac{dS}{dt} + \sum _i ^Q C_i
\label{eqn:F}
\end{equation}
where $\chi ^2$, $S$ and $C_i$ are all functions of the particle weights $\bmath{w} = (w_1, \cdot \cdot \cdot , w_N)$; $t$ is time; and $\mu$ and $\epsilon$ are positive parameters.  $Q$ is the number of additional constraints $C_i$.
The equations governing weight evolution over time come from maximising $F(\bmath{w})$ with respect to the particle weights ($\partial F / \partial w_i = 0,\;\; \forall i$) and rearranging terms to give equations of the form
\begin{equation}
	\frac{d}{dt} w_i = - \epsilon w_i G(\bmath{w}).
	\label{eqn:wtevoln}
\end{equation}
The overall rate of weight evolution is controlled by $\epsilon$.  The precise form of the function $G(\bmath{w})$ depends on the constraints $C_i$ and is illustrated later (equation \ref{eqn:gw}).  The process being applied to $\chi ^2$ is one of regularised, parameterised constrained extremisation.

The $\chi ^2$ term in $F$ arises from assuming that the probability of the model reproducing a single observation can be represented by a Gaussian distribution and then constructing a log likelihood function covering all observations. For $K$ different observables, we take $\chi ^2$ in the form
\begin{equation}
	\chi ^2 = \sum _k ^K \lambda _k \chi _k ^2
\label{eqn:chilambda}
\end{equation}
where $\lambda _k$ are small, positive parameters whose role is explained in section \ref{sec:params}.  Note that for statistical purposes, for example obtaining confidence limits or regions, equation \ref{eqn:chilambda} means that $\chi ^2$ is a linear combination of $\chi ^2$ terms. As such, it must not in general be associated with a $\chi ^2$ distribution but with a sum of gamma distributions.
\begin{equation}
	\chi ^2 _k = \sum _j ^{J_k} \Delta _{k,j} ^2
\end{equation}
and
\begin{equation}
	\Delta _{k,j} = \frac{y_{k,j}(\bmath{w}) -Y_{k,j}}{\sigma_{k,j}}
\end{equation}
where $Y_{k,j}$ is the measured value of observable $k$ at position $j$ with error $\sigma_{k,j}$, and $y_{k,j}(\bmath{w})$ is the model equivalent of $Y_{k,j}$.
\begin{equation}
	y_{k,j}(\bmath{w}) = \sum _i ^N w_i K_{k,j}(\bmath{r}_i, \bmath{v}_i) \delta(i \in k,j)
\end{equation}
where $K_{k,j}(\bmath{r}_i, \bmath{v}_i)$ is the kernel for observable $k$ evaluated at position $j$ for a particle with position $\bmath{r}_i$ and velocity $\bmath{v}_i$.  $\delta(i \in k,j)$ is a selection function and signifies that only particles which contribute to observable $k$ at position $j$ should be included in the calculation of $y_{k,j}$.  We have listed the kernels required for this paper in section \ref{sec:kernels}.

The entropy function $S$ in $F$ is
\begin{equation}
	S(\bmath{w}) = - \sum _i ^N w_i \ln (\frac{w_i}{m_i})
\end{equation}
where $m_i$ is taken as the initial value of a particle weight (in practice, we take $m_i = 1/N$).  $S$ is used for regularisation / smoothing purposes with the amount of regularisation being controlled by the parameter $\mu$. The derivative term $dS/dt$ indicates that over time we require the particle weights, and thus $S$, to be constant.  As demonstrated in \citet{Long2010}, the term behaves as the constraint $dS/dt = 0$.  The behaviour of the regularisation terms in the weight evolution equations is to move the particle weights towards $m_i / e$ rather than $m_i$.  \citet{Morganti2012} solve this problem by changing $S$ to
\begin{equation}
	S(\bmath{w}) = - \sum _i ^N w_i \left [ \ln (\frac{w_i}{m_i}) - 1 \right ].
\end{equation}
Modifying $S$ in this way causes no change to equation \ref{eqn:F}. Other regularisation schemes are possible, for example second derivative smoothing in integral space.  The challenge then is to produce a cost-effective implementation in a message passing computational environment.

The functions $C_i$ in $F$ are additional constraints to be included in the maximisation of $F$.  In this paper, we use only one such constraint which is that we require the model luminosity to match the luminosity ($L$) of the galaxy being modelled, that is $\sum Lw_i = L$, or more concisely $\sum w_i = 1$.  We therefore take
\begin{equation}
	C_1 = - \frac{\lambda_{\rmn{sum}}}{2}  \left ( \sum _i ^N  w_i - 1 \right ) ^2
\end{equation}
where $\lambda_{\rmn{sum}}$ is a positive parameter.

Given the definitions of $\chi ^2$, $S$ and $C_i$ and noting that for the purposes of this paper we do not use regularisation ($\mu = 0$, see section \ref{sec:Regularisation}), $G(\bmath{w})$ from equation \ref{eqn:wtevoln} can now be written
\begin{equation}
	G(\bmath{w}) = \sum _k ^K \lambda _k \frac{K_{k,ji}}{\sigma _{k,j}} \Delta _{k,j} + \lambda_{\rmn{sum}} \left ( \sum _ i ^N w_i - 1 \right )
\label{eqn:gw}
\end{equation}
where $K_{k,j}(\bmath{r}_i, \bmath{v}_i)$ has been abbreviated to $K_{k,ji}$ and we have assumed that, for all observables, 1 particle contributes only at 1 position $j$.

Finally, model observables (and thus particle weights) are subject to noise as the numbers of particles contributing to the observables vary.  This noise is suppressed by replacing $\Delta _{k,j}$ in $G(\bmath{w})$ by an exponentially smoothed version $\tilde{\Delta}_{k,j}$ given by
\begin{equation}
  \frac{d}{dt} \tilde{\Delta}_{k,j} = \alpha ( \Delta _{k,j} - \tilde{\Delta}_{k,j} )
\end{equation}
where $\alpha$ is a small positive parameter.  The smoothed $\Delta _{k,j}$ can be used to calculate a smoothed version $\tilde{y} _{k,j}$ of the model observable,
\begin{equation}
	\tilde{y} _{k,j} = Y_{k,j} + \sigma _{k,j} \tilde{\Delta}_{k,j}.
\end{equation}
Assuming the calculation suggested in \citet{Syer1996} is followed, the effect of exponential smoothing is as if the number of particles has been boosted by two or three orders of magnitude.

\subsection{Regularisation}\label{sec:Regularisation}
Given that the number of particles typically exceeds the number of observations by several orders of magnitude, 
determining the particle weights using the M2M method is an ill-posed problem with potentially many solutions.
Regularisation ($S$) is introduced to avoid overfitting the data and to reduce the number of solutions to those that satisfy the regularisation condition.  In the M2M case, regularisation behaves to smooth the M2M model either at a global level (entropy based functions) or at a local level (second derivative smoothing).  The amount of regularisation is controlled by the parameter $\mu$ which is used to achieve some desired balance between 
fitting the observed data and delivering a smooth solution.  Quite what that balance should be is up to the
individual researcher to determine.  Entropy based regularisation introduces another consideration in that it
prevents the particle weights ($w_i$) moving too far from their priors ($m_i$) and thus the choice of priors has a major influence on the model solution.

For the investigation in this paper (determining the bar angle and pattern speed), we are using the M2M method 
as a `black box' and are more concerned with differences between models, as represented by the end of run model $\chi^2$ values, rather than the detailed solutions represented by individual models.  The bar angle and the pattern speed are global properties of the galaxy being modelled and we do not wish the values we determine to be influenced by the behaviour of any regularisation condition.  We thus choose deliberately to model with no regularisation.  This approach is consistent with that in \citet{Long2012} where we successfully used the M2M method as a `black box' to determine a different global property (the mass-to-light ratio) of the galaxies being modelled.

As will be seen in the rest of this paper, the values of the bar angle and pattern speed we achieved are consistent with \citet{Shen2010} and we meet our objectives without employing regularisation.  We will comment further on regularisation in our conclusions in section \ref{sec:conclusions}.

\section{Modelling with Milky Way observables}\label{sec:milkyway}
The two main issues to be addressed here are 
\begin{enumerate}
\item modelling with Galactic field based observables where the fields are non-contiguous and specified in terms of Galactic longitude and latitude and can not be modelled as parallel projections, and
\item using a rotating frame (representing the bar) for orbit integration.
\end{enumerate}
Both these issues are straightforward to resolve in a M2M context.

\subsection{Observables}\label{sec:observables}
\begin{figure}
		\centering
		\includegraphics[width=75mm]{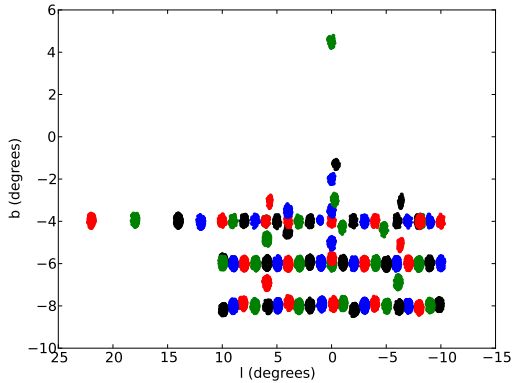}\\
		\caption[BRAVA fields]{The available BRAVA fields. There is no significance to the colour coding other than to distinguish adjacent or overlapping fields.}
\label{fig:bravafields}
\end{figure}

We take the measured kinematic data for our models from the publicly available BRAVA data\footnote{http://brava.astro.ucla.edu/data.htm} (downloaded in December 2011).  The M2M observables thus comprise mean radial velocity and velocity dispersion squared in a number of Galactic fields specified by Galactic longitude and latitude $(l,b)$.  For consistency with \citet{Shen2010}, we do not use the individual stellar velocity measurements within each field as M2M constraints.  We do however include the $b=-6^{\circ}$ BRAVA measurements which were not available to \citet{Shen2010}.  Figure \ref{fig:bravafields} shows the individual fields.  For modelling purposes, we define the boundaries of each field as the minimum and maximum longitude and latitude. In total, $86$ $(l,b)$ fields where downloaded of which $78$ distinct fields are used in our models.  The difference is due to overlaps occuring in $15$ of the fields with $(l,b)$ values of $(-8^{\circ}, -4^{\circ})$, $(4^{\circ}, -5^{\circ})$, $(0^{\circ}, -3.5^{\circ})$, $(0^{\circ}, -6^{\circ})$, $(10^{\circ}, -6^{\circ})$ and $(6^{\circ}, -4^{\circ})$.  After visual examination of the degree of overlap, 8 of the overlapping fields were arbitrarily removed from the modelling.  Given the low number, the fact that the spread of measurement points was not impacted, and the nature of the exercise (results consistency check only), further investigation of the overlapping fields was considered unnecessary.

`Measured' luminosity observables are calculated directly from the \citet{Shen2010} particle data.  Luminosity density is taken to be axisymmetric (with z-axis, the symmetry axis) and is formed by binning particles in $(R,z)$ where $R$ is the equatorial plane projected radius.  All particles at this stage are assumed to have equal luminosity. The observed values are calculated out to $R = 13 \: \rm{kpc}$.  For a projected luminosity observable, we use the fractional field luminosity where the fields match the BRAVA $(l, b)$ fields, and construct it by counting particles accordingly.  To take into account that the measured observables are determined by binning particle data (and are therefore subject to noise), and to allow the M2M models the opportunity to retain the bar shape (which is not axisymmetric), we set the error terms for the observables to a high relative error of $20\%$.  

It is worth noting that, at any one time, only $\approx 2.5 \times 10^4$ particles (out of $\approx10^6$ particles in total) are present in the BRAVA fields and contributing directly to the model observable calculations.  For binning particles into fields, we implement a scheme which first eliminates quickly particles which are not in any field before performing a more detailed field selection.

The units we use within the M2M models in this paper are kilo-parsecs for length, $10^7$ years for time, and mass in units of the solar mass $M_{\odot}$.

\subsection{Kernels}\label{sec:kernels}
We list the kernels used for calculating model observables. In the equations below, $L$ is the total luminosity of the model, $V_j$ is the volume of observable bin $j$, $v_{\rm{radial},i}$ is the radial velocity (with respect to the Sun's position) of particle $i$, and $L_j$ is the fractional luminosity of field $j$.
\begin{enumerate}
\item luminosity density
\begin{equation}
	K_{ji} = \frac{L}{V_j}
\end{equation}

\item fractional field luminosity
\begin{equation}
	K_{ji} = 1
\end{equation}

\item 1st radial velocity moment
\begin{equation}
	K_{ji} = \frac{v_{\rm{radial},i}}{L_j}
\end{equation}

\item 2nd radial velocity moment
\begin{equation}
	K_{ji} = \frac{v_{\rm{radial},i} ^2}{L_j}
\end{equation}

\end{enumerate}

We calculate the model radial velocity dispersion squared directly from the 1st and 2nd velocity moments using
\begin{equation}
	\sigma _j ^2 = \overline{v ^2} _j - (\overline{v}_j)^2
\end{equation}
where, for bin $j$, $\sigma _j$ is the dispersion, $\overline{v ^2} _j$ is the 2nd moment, and $\overline{v} _j$ is the  1st moment.

\subsection{Gravitational Potential}\label{sec:potentials}
As in \citet{Shen2010}, the gravitational potential comprises 2 components, a dark matter halo given by
\begin{equation}
	\phi(r) = \frac{1}{2} V_c ^2 \ln \left ( 1 + \frac{r^2}{R_c ^2} \right )
\end{equation}
where $V_c = 248 \: \rm{km/s}$ and $R_c = 15.2 \: \rm{kpc}$, and a potential derived from the luminous matter particles.

We do not use self-gravitation of the particles as in \citet{Shen2010} and comment on this later in our conclusions (section \ref{sec:conclusions}).  Instead, we calculate the luminous matter potential and its associated accelerations once from the particle positions using direct summation and hold the values in interpolation tables. The tables use cylindrical polar coordinates with a pseudo-logarithmic radial coordinate. We use tri-linear interpolation to determine the function values as required at specific positions.  We initially used the self-consistent field method of \citet{Hernquist1992} to determine the potential from the particles but stopped when we discovered that, under certain circumstances, densities calculated via the method were negative. Why this occurred is currently unresolved.  

Direct summation is straightforward to parallelise using a graphics processor unit (GPU) and we currently achieve a four-fold increase in performance over a single computer processor.  With experience, we would expect this increase to become greater.

The total mass of the particle model is $4.25 \times 10^{10} M_{\odot}$.  For implementation reasons (the particle model is a mass model and the M2M theory in section \ref{sec:m2m} is framed in terms of luminosity), the M2M models assume a luminous matter mass-to-light ratio of $1$.

\subsection{Particle initial conditions}\label{sec:particleics}
\begin{figure*}
		\centering
		\includegraphics[width=56mm]{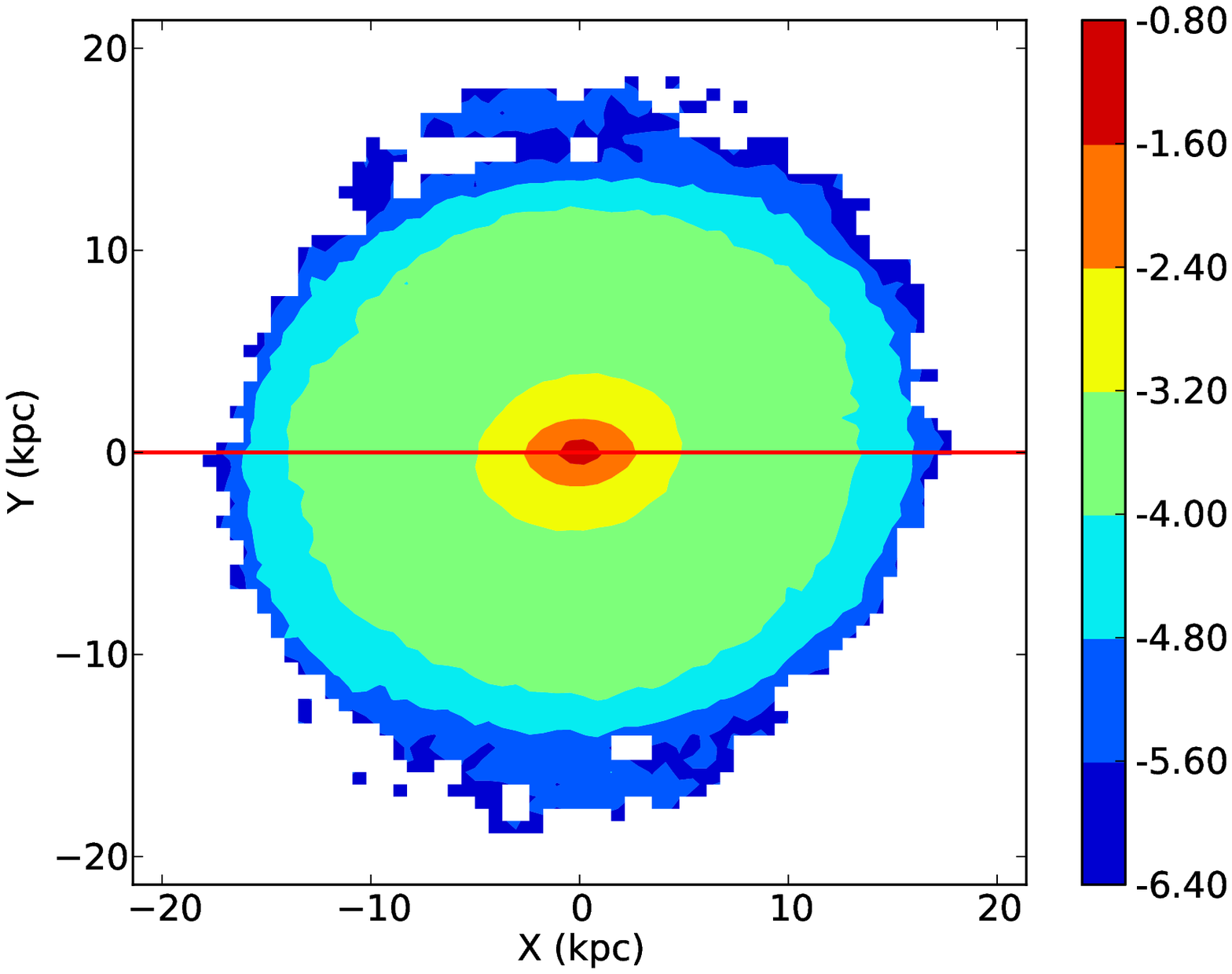}
		\includegraphics[width=56mm]{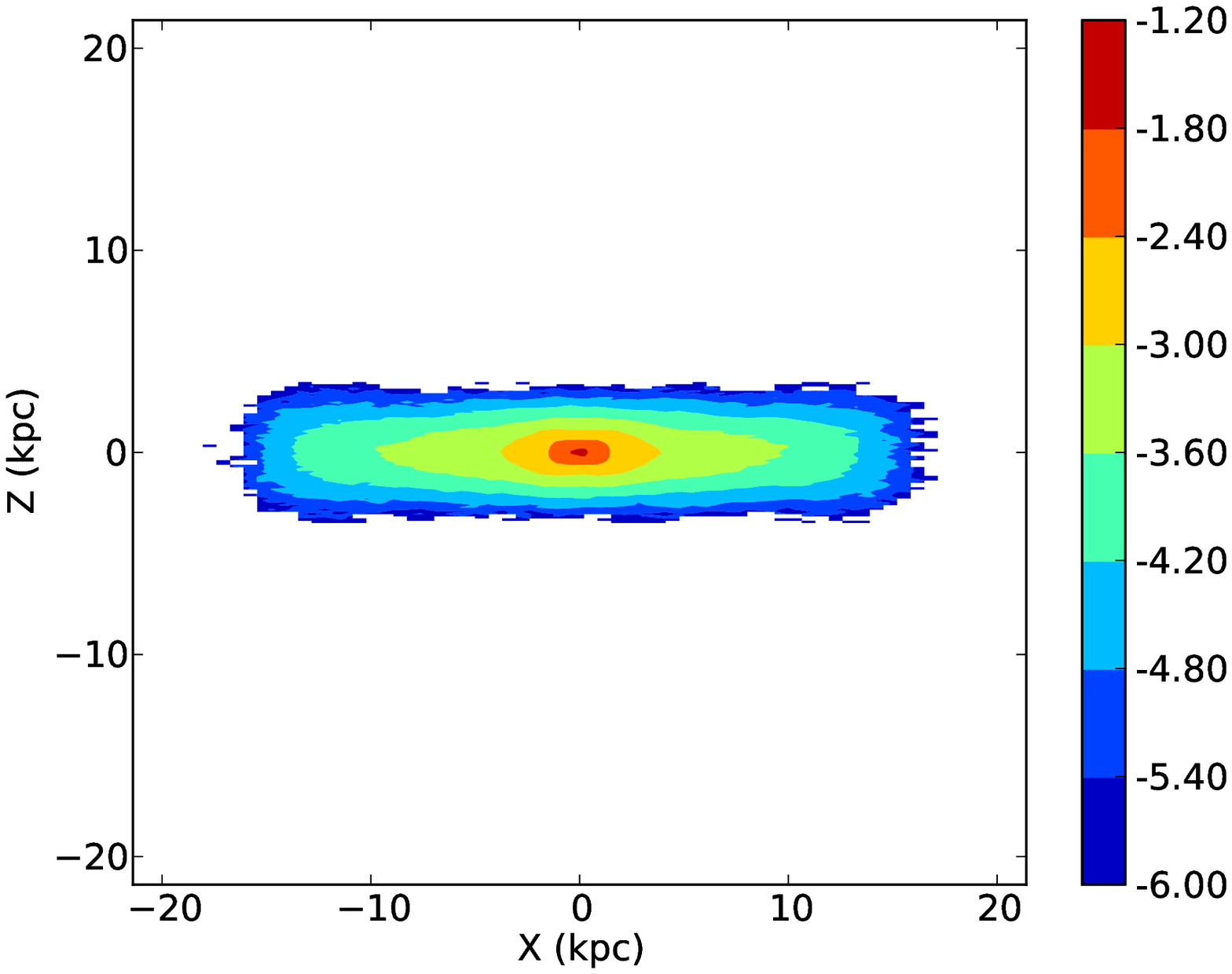}
		\includegraphics[width=56mm]{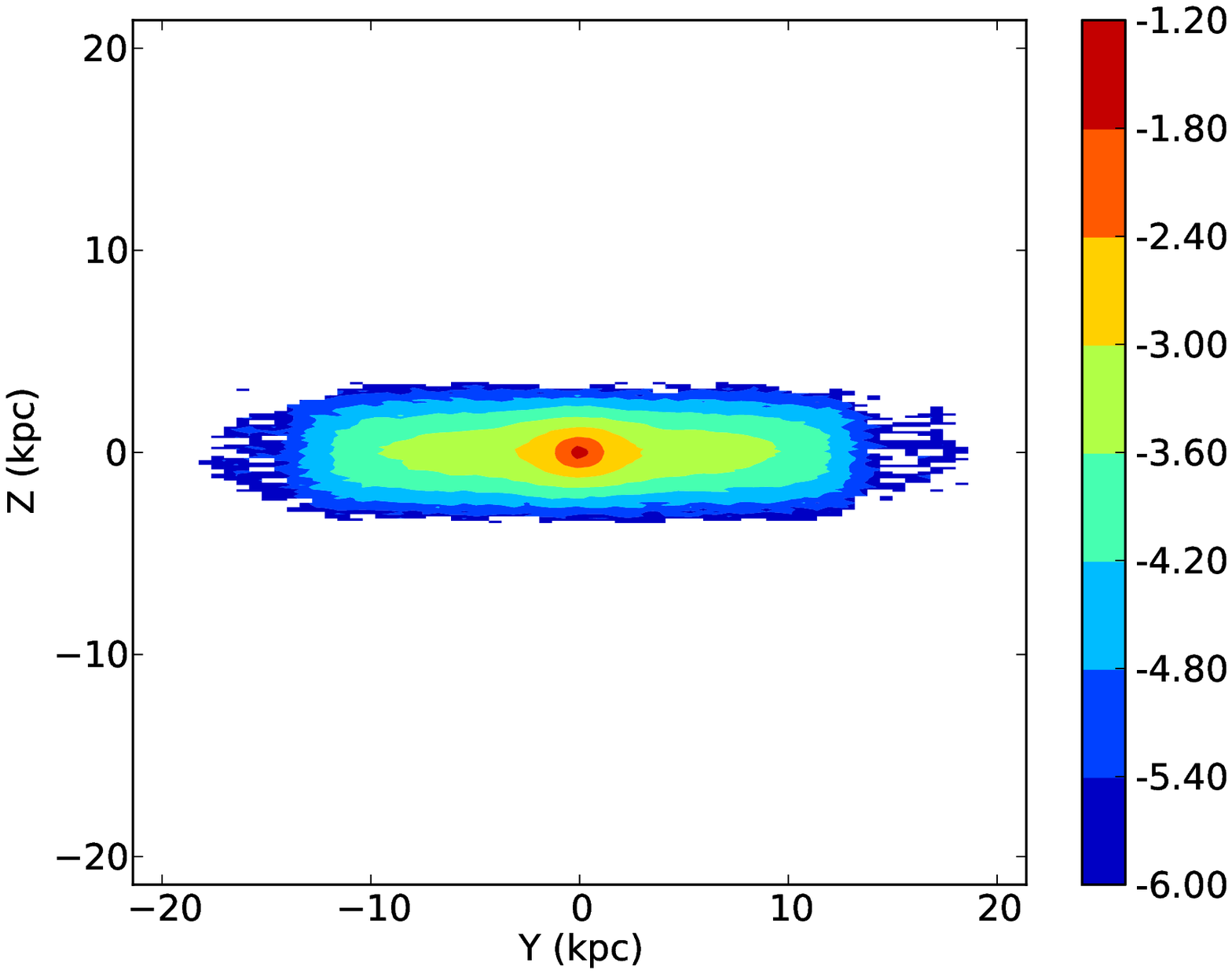}
		\caption[Initial particle positions]{Logarithmic fractional projected particle counts showing the initial spatial distribution of the particles. The bar lies along the x-axis (indicated by the red line in the left hand panel).  During modelling, the particles are first rotated to the desired bar angle before orbit integration starts.}
\label{fig:particles}
\end{figure*}

The initial positions and velocities for the particles are the end of run values taken from \citet{Shen2010}.  Plots of the projected initial positions of the particles are shown in Figure \ref{fig:particles}.  In these plots, the bar lies along the x-axis.  During M2M modelling, the particles are first rotated to the desired bar angle before orbit integration starts.

The initial weights of the particles are set as $1/N$ where $N$ is the number of particles.

\subsection{Rotating frame kinematics}
It is convenient to think of the M2M model in terms of 3 coordinate frames (inertial, rotating and constraints)  with a common origin at the Galactic centre.  

\begin{figure}
		\centering
		\includegraphics[width=56mm]{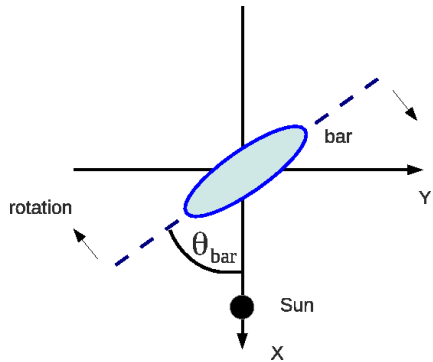}
		\caption[Equatorial plane]{Schematic showing the inertial equatorial plane of the M2M model. $\theta _{\rm{bar}}$ is the bar angle and has a negative value when mimicking the Galactic bar.  Given the sense of rotation of the bar is clockwise, the pattern speed ($\Omega _{\rm{p}}$) is also negative. Positive Galactic longitudes are on the left of the schematic.}
\label{fig:eqplane}
\end{figure}

The \textit{inertial} frame represents the overall model with the Sun stationary on the positive x-axis  at $8.5 \: \rm{kpc}$ from the Galactic centre.  The x-y plane is the equivalent of the Galactic equatorial plane and is shown diagrammatically in Figure \ref{fig:eqplane}.  The \citet{Shen2010} particle data has the bar positioned along the x-axis initially with a clockwise sense of rotation.  For modelling purposes, the bar is first positioned at some desired bar angle prior to orbit integration starting.

The \textit{rotating} frame rotates at a given pattern speed ($\Omega _{\rm{p}}$), fixed per modelling run, and contains the bar (and indeed all the particles).  Orbit integration takes place in the rotating frame and uses a leapfrog scheme based on \citet{Pfenniger1993}.  We use the leapfrog scheme in `drift-kick-drift' form with a constant time step of $2 \times 10^{-3}$ time units.  Jacobi's integral is conserved (see \citet{BT2008}) to a relative accuracy of $\approx 10^{-2}$.  This value is the mean maximum relative accuracy taken across all particles for the duration of a modelling run.  Given we are using interpolation tables to hold the potential and its accelerations, and are only seeking consistency with the results in \citet{Shen2010}, the value is satisfactory for our purposes.  We orbit the particles for $125$ time units.  For $\Omega _{\rm{p}} = -40 \: \rm{km/s/kpc}$, this equates to $\approx 8$ full revolutions of the bar.

Observable constraints must only be applied with the bar at the desired bar angle to the Sun to Galactic centre line.  Conceptually, the \textit{constraints} application frame must therefore be rotated as the bar rotates to keep the bar angle at the desired value.  This is the equivalent of either rotating the Sun's position or reversing the bar's rotation for the purposes of applying the constraints. In practice, to avoid manipulating the constraints or their fields, we choose the latter.

\subsection{Parameter setting}\label{sec:params}
In Table \ref{tab:params} we list all the parameters identified in section \ref{sec:theory} and the values we use in our M2M models.  The values are the same for all models.  The rationale for the $\lambda _k$ parameters in equation \ref{eqn:chilambda} is recorded in \citet{Long2010} and \citet{Long2012} and is not repeated here.  These parameters have only been set to an order of magnitude.  No fine tuning, or sensitivity analysis, has been undertaken. $\lambda _{\rm{sum}}$ has been determined by experimentation to ensure $\sum w_i \; = \; 1.00$ for all models.

\begin{table}
	\centering
	\caption{M2M parameters}
	\label{tab:params}
	\begin{tabular}{ccc}
		\hline
		Parameter & Value & Related constraint \\
		\hline
		$\epsilon$ & $2.5 \times 10^{-3}$ & \\
		$\alpha$   & $5.0 \times 10^{-2}$ & \\
		$\mu$      & $0$                  & \\
		$\lambda _{\rm{sum}}$ & $10^4$    & sum of particle weights \\
		$\lambda _{\rm{LD}}$  & $10^{-3}$ & luminosity density \\
		$\lambda _{\rm{FL}}$  & $10^{-3}$ & fractional field luminosity \\
		$\lambda _{\rm{V}}$   & $10^{-4}$ & radial velocity \\
		$\lambda _{\rm{VD}}$  & $10^{-4}$ & radial velocity dispersion \\
		\hline
	\end{tabular}
	
\medskip
The M2M parameters and their values.  The same values are used in all modelling runs.  The rationale for the observable $\lambda$ parameters is recorded in \citet{Long2010} and \citet{Long2012}.  $\lambda _{\rm{sum}}$ is set by experimentation to ensure $\sum w_i \; = \; 1.00$.
\end{table}

\section{Results}\label{sec:results}
\begin{figure*}
		\centering
		\includegraphics[width=75mm]{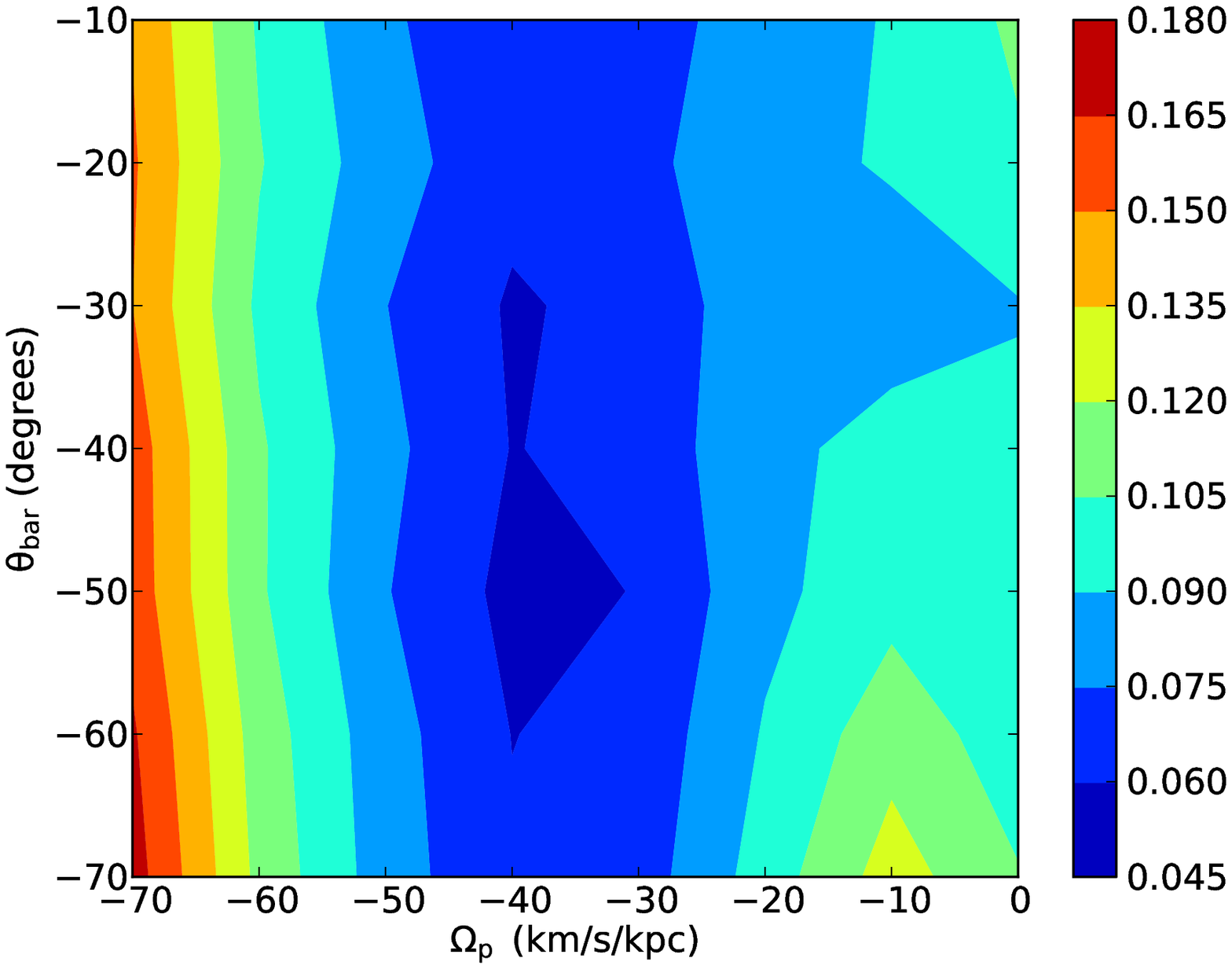}
		\includegraphics[width=75mm]{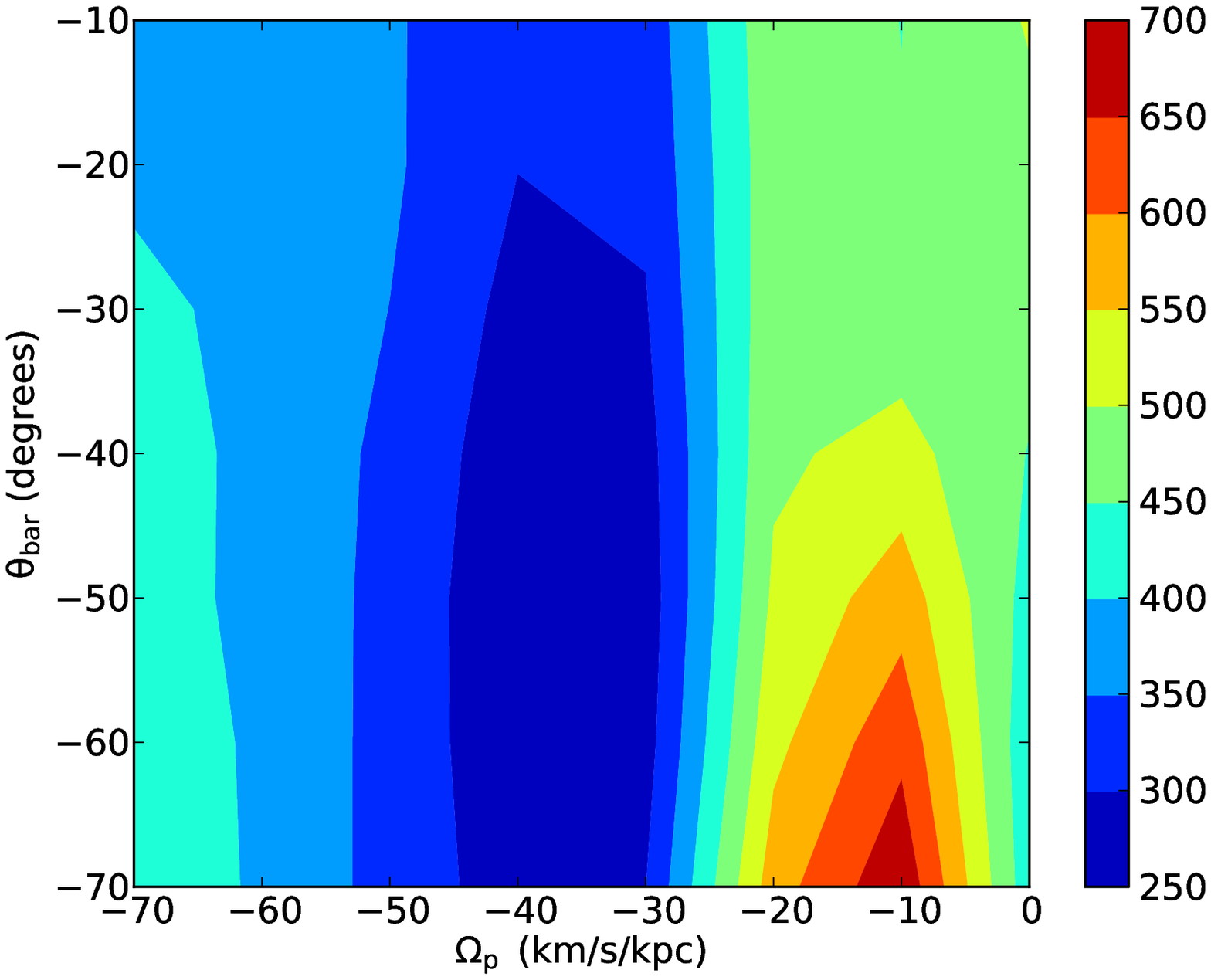}\\
		\includegraphics[width=75mm]{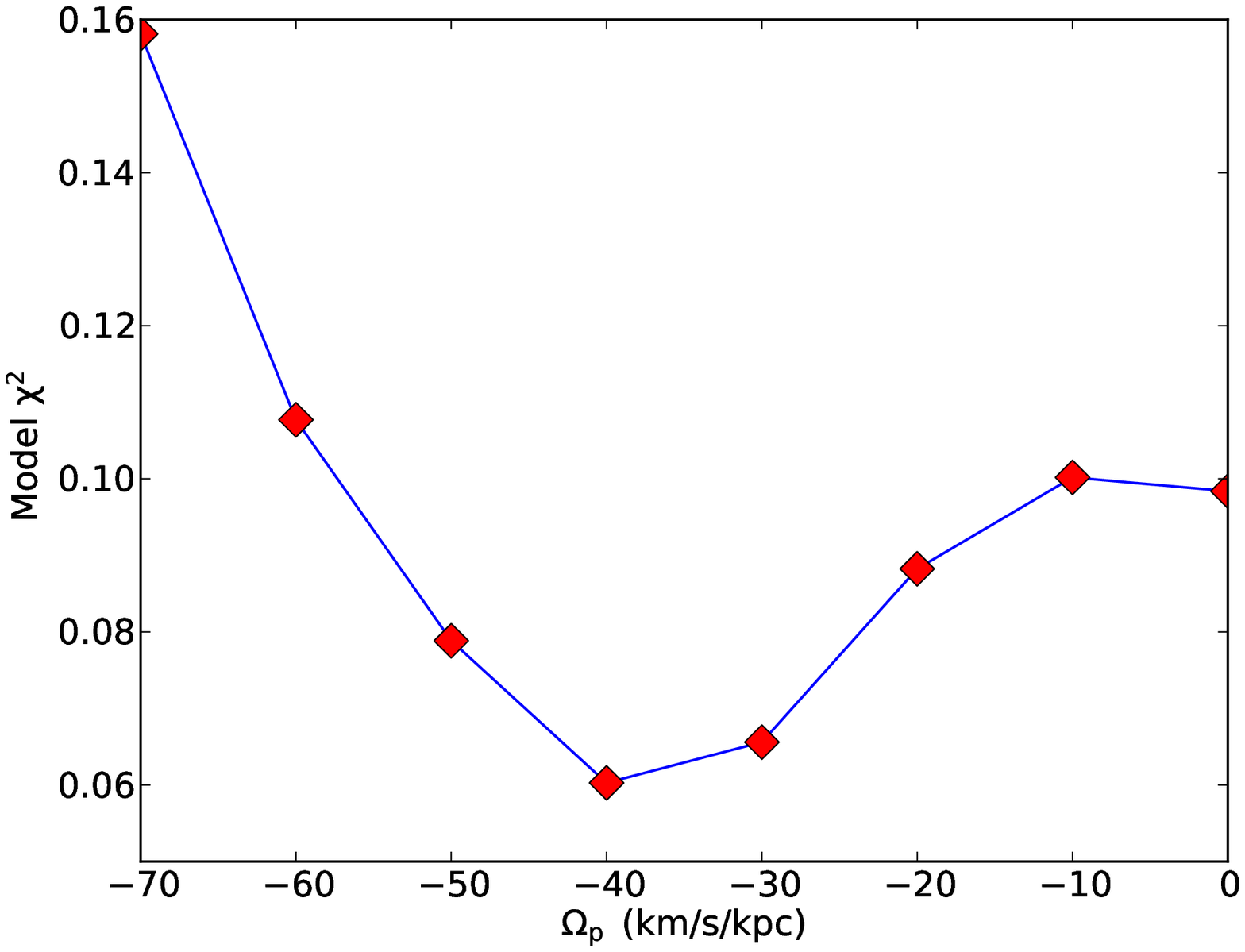}
		\includegraphics[width=75mm]{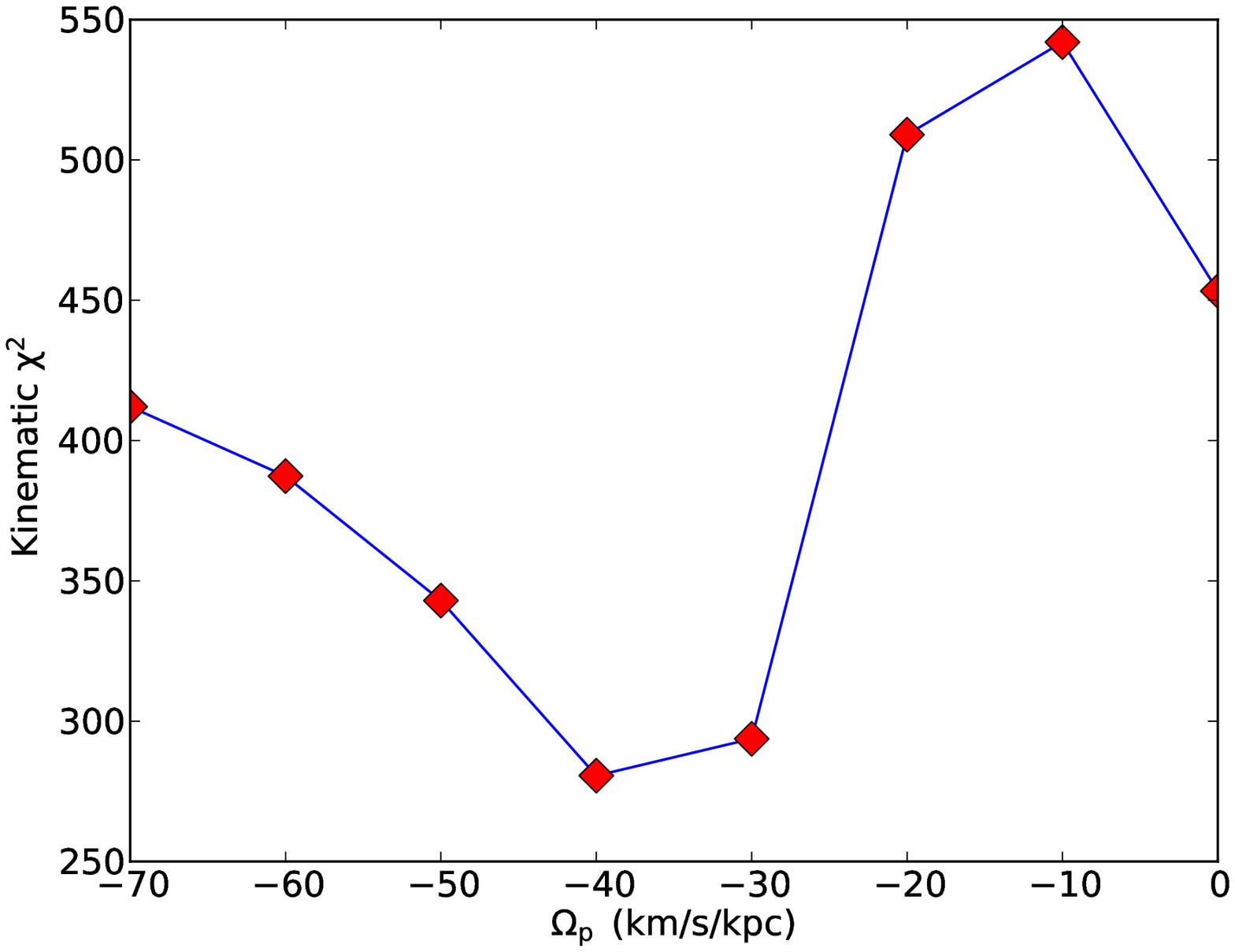}\\
		\includegraphics[width=75mm]{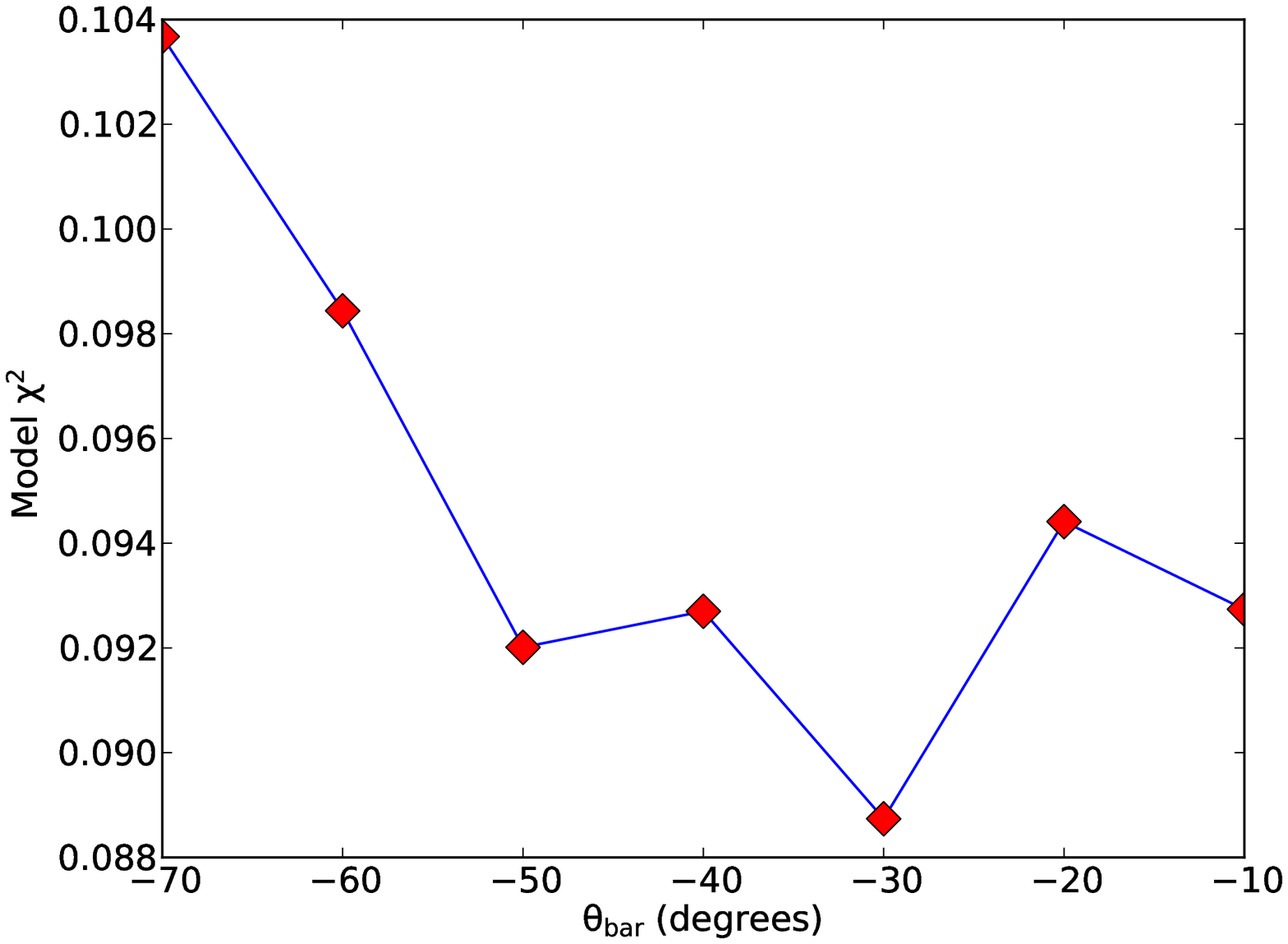}
		\includegraphics[width=75mm]{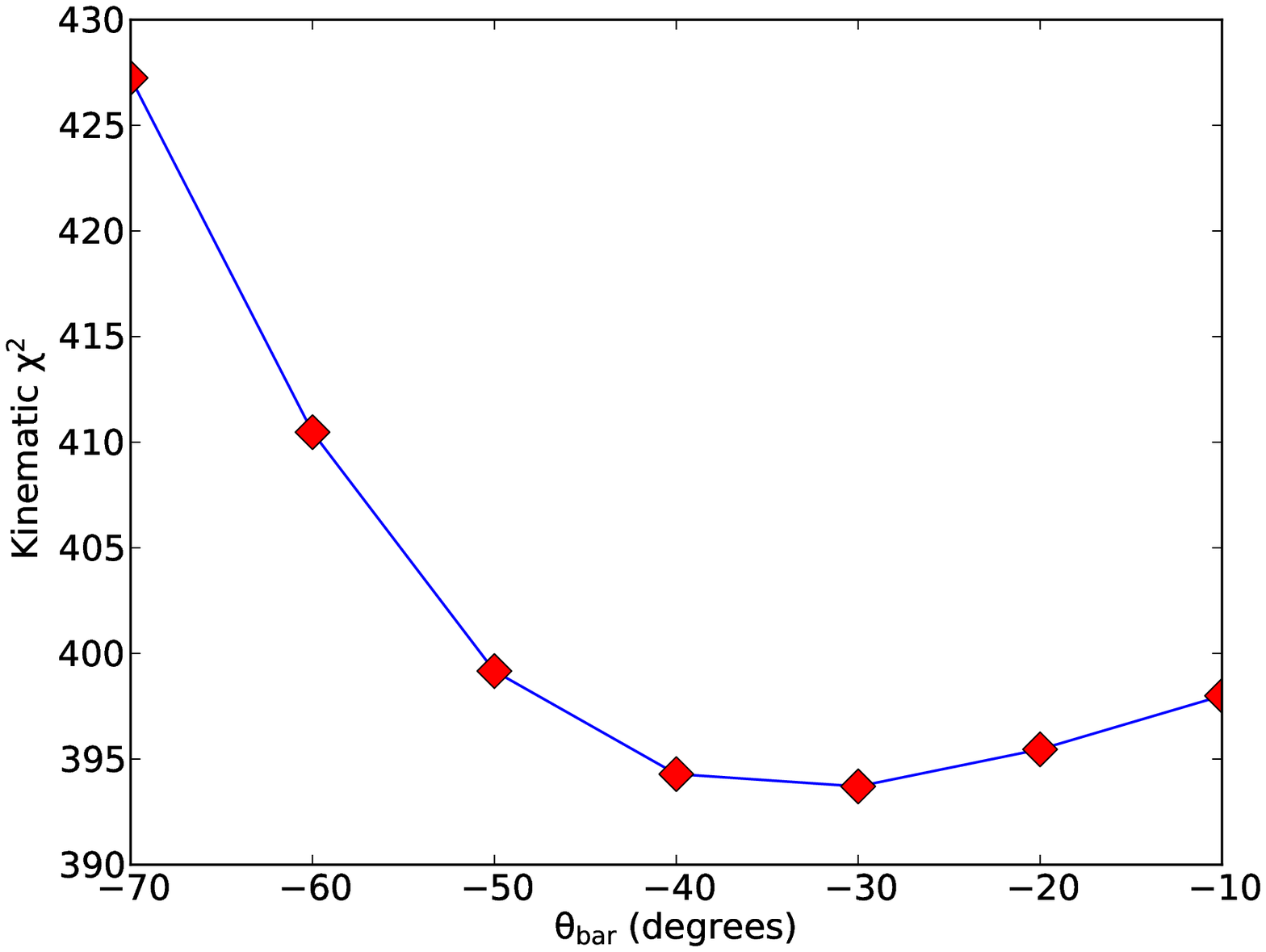}\\
		\caption[$\chi ^2$ plots]{$\chi ^2$ surface plots and marginalised plots for the M2M models using both density and kinematic constraints (left hand column), and for the kinematic constraints only (right hand column). In the left hand column, $\chi ^2$ is calculated using equation \ref{eqn:chilambda} involving the $\lambda _k$ parameters. No $\lambda _k$ values are used in calculating the kinematic $\chi ^2$ in the right hand column. In the marginalised plots, the data points are shown in red; the blue lines are for visual guidance only.  $\Omega _{\rm{p}}$ is the pattern speed and $\theta _{\rm{bar}}$, the bar angle.}
\label{fig:chi2plots}
\end{figure*}

By varying the bar angle $\theta _{\rm{bar}}$ between $-10^{\circ}$ and $-70^{\circ}$ with a $10^{\circ}$ interval, and the pattern speed $\Omega _{\rm{p}}$ between $0 \: \rm{km/s/kpc}$ and $-70 \: \rm{km/s/kpc}$ with a $10 \: \rm{km/s/kpc}$ interval, we construct a set of 56 M2M models.  Using these models, we create a model $\chi ^2$ surface where $\chi ^2$ is as in equation \ref{eqn:chilambda} and involves the $\lambda _k$ values. By marginalising $\chi^2$ over $\theta _{\rm{bar}}$ and $\Omega _{\rm{p}}$ individually, we obtain the best fit values of $\theta _{\rm{bar}} \approx -30^{\circ}$ and $\Omega _{\rm{p}} \approx -40 \: \rm{km/s/kpc}$.  These values are consistent with the values \citet{Shen2010} determined for their N-body model.  If we now take only the kinematic components from $\chi ^2$, without their corresponding $\lambda _k$ values, and repeat the marginalisation exercise, we arrive at the same values for $\theta _{\rm{bar}}$ and $\Omega _{\rm{p}}$.  The $\chi ^2$ surfaces and marginalised plots are shown in Figure \ref{fig:chi2plots}.  It is clear from the figure that the bar angle is only weakly determined by the kinematic observations alone, and is better determined when the luminosity constraints are included.

\begin{figure*}
		\centering
		\includegraphics[width=170mm, height=100mm]{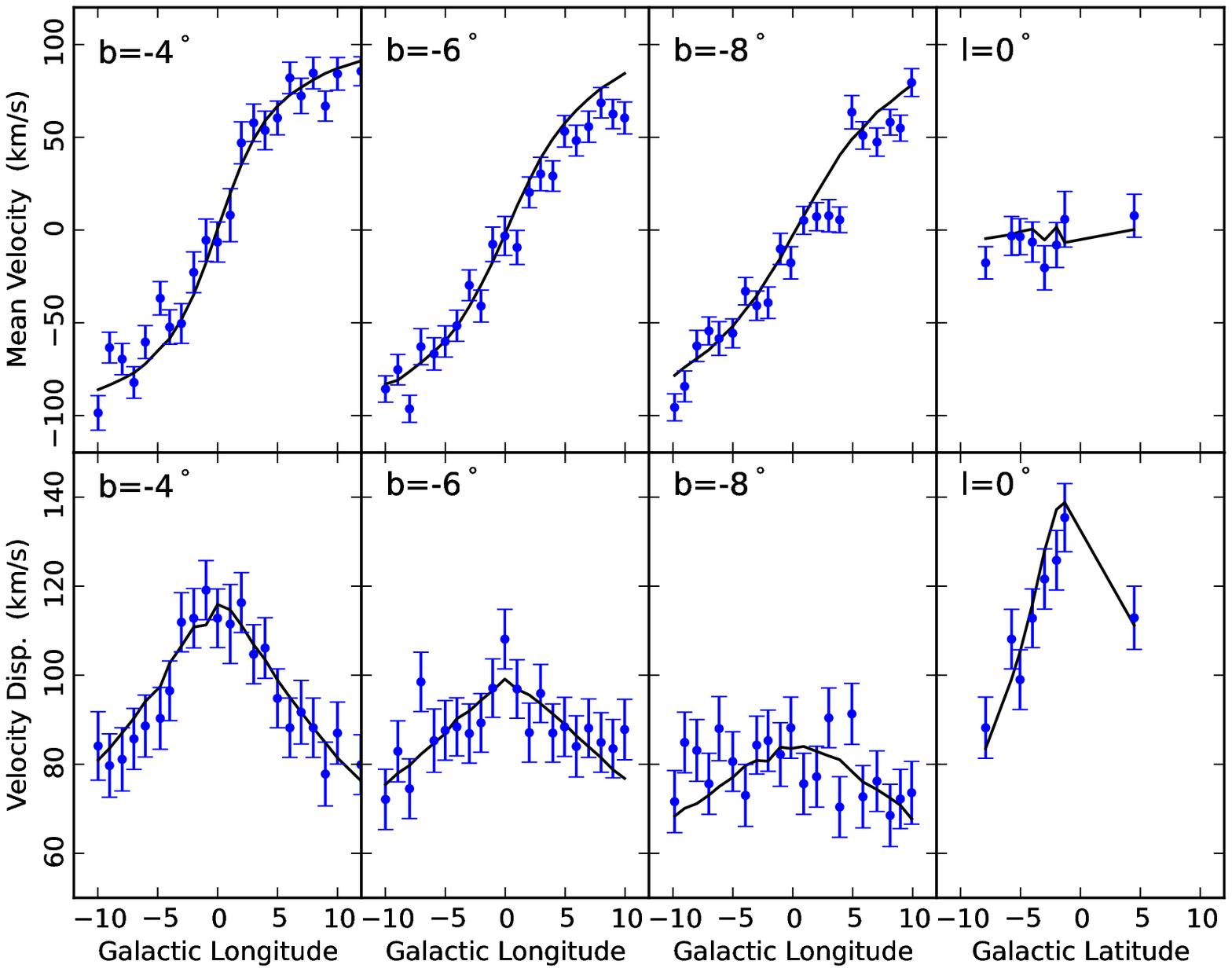}
		\caption[Bar structure]{Reproduction by the best fit M2M model ($\theta _{\rm{bar}} \approx -30^{\circ}$ and $\Omega _{\rm{p}} \approx -40 \: \rm{km/s/kpc}$) of the observed velocities and velocity dispersions along the minor axis ($l=0^{\circ}$) and by latitude ($b=-4^{\circ}, -6^{\circ}, -8^{\circ}$).  The black solid lines represent the model and the observed values are in blue.  See Figure 2 in \citet{Shen2010} for a comparison with the N-body results. Note that positive Galactic longitude is to the right in the plots to aid that comparison.}
\label{fig:bygalcoords}
\end{figure*}

\begin{figure*}
		\centering
		\includegraphics[width=170mm, height=100mm]{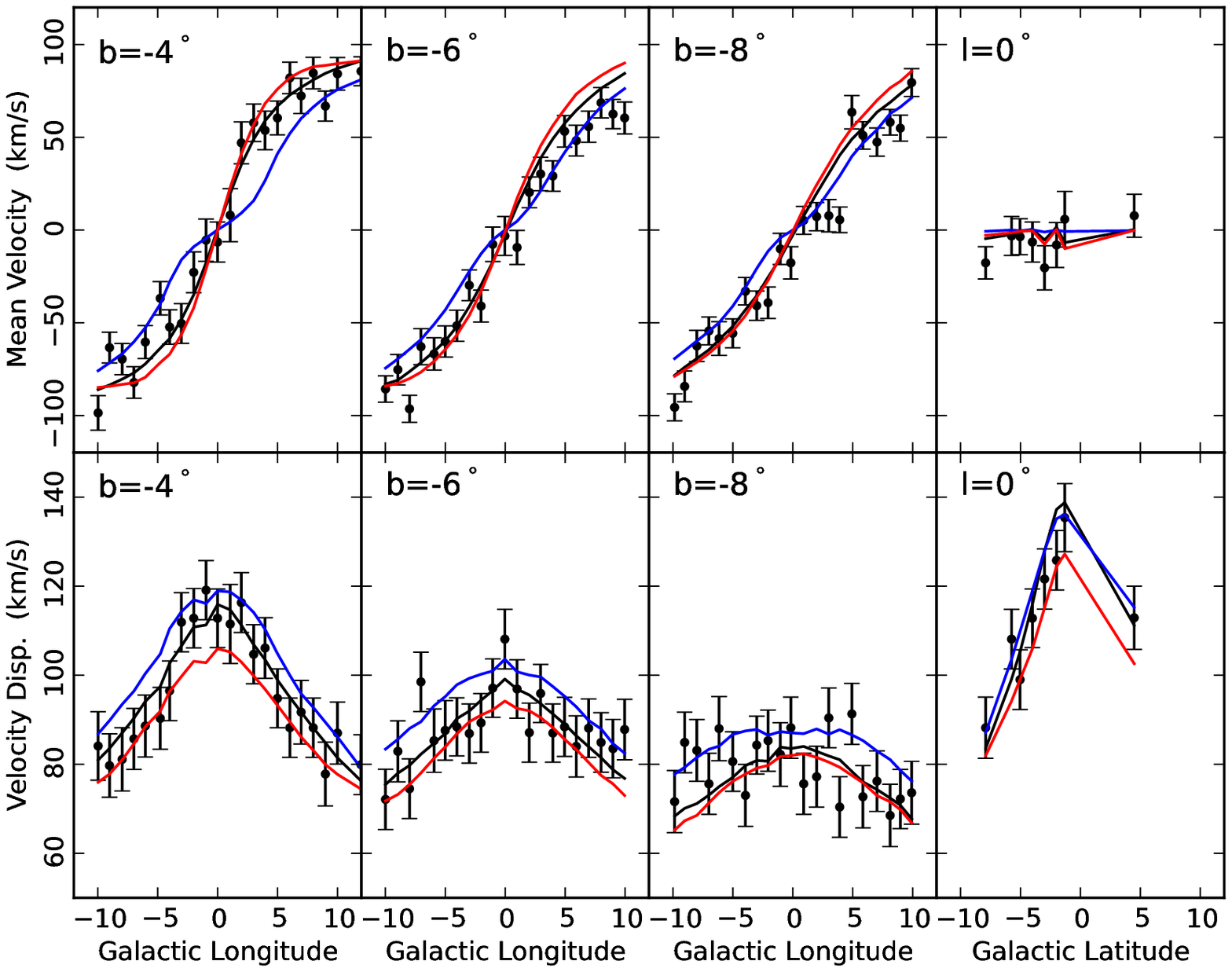}
		\caption[Bar structure]{Reproduction by the best fit M2M model ($\theta _{\rm{bar}} \approx -30^{\circ}$ and $\Omega _{\rm{p}} \approx -40 \: \rm{km/s/kpc}$) of the observed velocities and velocity dispersions along the minor axis ($l=0^{\circ}$) and by latitude ($b=-4^{\circ}, -6^{\circ}, -8^{\circ}$).  The black solid lines represent the best fit model and the black points with error bars, the observed values.  With the bar angle at the best fit value, the blue and red lines show the observable reproduction for patterns speeds of $\Omega _{\rm{p}} = 0 \: \rm{km/s/kpc}$ and $\Omega _{\rm{p}} = -70 \: \rm{km/s/kpc}$ respectively.}
\label{fig:bygalcoordsba}
\end{figure*}

The best fit model ($\theta _{\rm{bar}} \approx -30^{\circ}$ and $\Omega _{\rm{p}} \approx -40 \: \rm{km/s/kpc}$) reproductions of the BRAVA field mean radial velocities and velocity dispersions are given in Figure \ref{fig:bygalcoords}.  The equivalent figure in \citet{Shen2010} is Figure 2.  The model reproductions of the kinematics are quite smooth and do not suggest that regularisation should have been used. The $\chi ^2$ per degrees of freedom values for the best fit model are $2.55$ for the radial velocities and $1.09$ for the velocity dispersions.  It may be seen from Figure \ref{fig:bygalcoords} (see the $b=-6^{\circ}$ and $b=-8^{\circ}$ panels) that the scatter in the field radial velocity values and the size of the error bars is causing the M2M models to be unable to reproduce all the measured velocity values, resulting in higher $\chi ^2$ values not just in the best fit model but in all models.  The $\chi ^2$ per degrees of freedom values for luminosity density and fractional field luminosity are both low at $\approx 0.1$.  This is almost certainly due to using high relative errors ($20\%$) as noted in section \ref{sec:observables}.  However, given the consistency objective in section \ref{sec:introduction} had been met, no experimentation took place with lower error values.  For comparison with Figure \ref{fig:bygalcoords}, Figure \ref{fig:bygalcoordsba} shows the variation in observable reproduction for pattern speeds either side of the best fit value, with the bar angle at its best fit value.

\begin{figure}
		\centering
		\includegraphics[width=75mm]{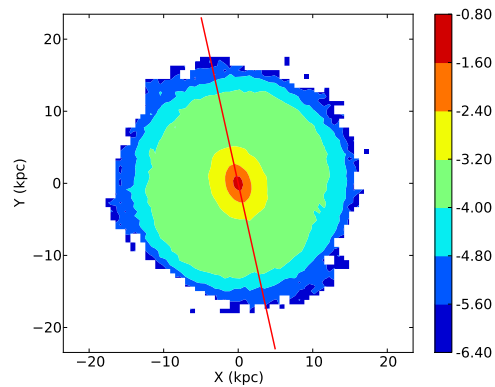}
		\caption[Bar structure]{Projection onto the equatorial plane of the logarithmic particle weight distribution for our best fit model ($\theta _{\rm{bar}} \approx -30^{\circ}$ and $\Omega _{\rm{p}} \approx -40 \: \rm{km/s/kpc}$). The bar structure (see Figure \ref{fig:particles}) is maintained during orbit integration and is in the expected position at end of run (indicated by the red line).}
\label{fig:bar}
\end{figure}

Notwithstanding that an axisymmetric distribution has been used in constructing the luminosity density constraint, the bar structure has been maintained during orbit integration as can be seen from Figure \ref{fig:bar}.  Weight convergence for our best fit model is high with $99\%$ of particles having converged weights (the definition of weight convergence is given in \citealt{Long2010}).  The final weights of the best fit model particles remain close to their initial values ($1.02 \times 10^{-6}$) and are in the range $5.9 \times 10^{-7} < w_i < 2.8 \times 10^{-6}$.  The variation in weight range is most noticeable along the pattern speed axis where the minimum of the range is at a maximum for the best fit pattern speed $\Omega _{\rm{p}} \approx -40 \: \rm{km/s/kpc}$ and decreases on either side to $\approx 10^{-7}$ at 
$-70 \: \rm{km/s/kpc}$ and $\approx 10^{-8}$ at $0 \: \rm{km/s/kpc}$.  As a final comment, the narrow weight range of the best fit model is not unexpected given the way the luminosity data was constructed and the use of the BRAVA data in the N-body model.  The variation with pattern speed we take as a positive indicator that the M2M method used as a `black box' is able to discriminate successfully between top level parameter values.

As a guide to the computing resources required, one M2M modelling run, using $32$ processing cores, takes $\approx 60$ minutes to complete.  Assuming the M2M models in this exercise are run sequentially, the full set takes $56$ elapsed hours to complete.  We did not encounter the data packet fragmentation issue noted in \citet{Long2012} as the number of measurement points (equal to the number of fields) is low. Whether other M2M modellers will hit the fragmentation issue depends on their M2M implementation and its use of the Message Passing Interface (MPI), the variant of MPI used, and the network configuration of the hosting computer system, as well as the number of observable measurement points for the stellar system being modelled.

\section{Discussion on combined N-body and M2M modelling}\label{sec:discussion}

We now consider further combining N-body and M2M modelling.  Our initial thoughts on starting the current exercise were that we could forsee a combined modelling process with repeating interleaved N-body and M2M phases with each phase seeking to refine the preceding phase.  This paper represents the first step in developing that process in that it seeks to ensure that M2M modelling delivers results consistent with an established N-body stellar dynamical model.  However the interleaving approach is not the only way forward and we now describe some alternatives.

There appear to be four main options:
\begin{enumerate}
\item keep N-body and M2M modelling completely separate so that they can be used to perform an independent check on each other,
\item loosely couple the 2 methods so that the output from an N-body model can be fed into a M2M model (with the       ability to iterate between model types),
\item absorb N-body modelling into M2M modelling, and
\item absorb M2M modelling into N-body modelling.
\end{enumerate}
These options are not necessarily disjoint.  Practically, in a software engineering sense, the last three options may not actually turn out to be that dissimilar.  For a given modelling scenario it may be appropriate to reuse one model type with multiple models of the second type, as in the current paper where one N-body model was used with many M2M models.  Clearly the N-body model could have been re-established for every M2M model but that would not offer a computationally cost effective solution. In other words, even in a combined model, there is a case for keeping the N-body and M2M modelling phases separate. 

The one-to-many scenario is appropriate when the reason for modelling is the determination of some global property or attribute of a stellar system, for example, the mass-to-light ratio or the bar pattern speed.  In this case, individual models are less important than the comparison between the models.  However, there are cases where the individual models take precedence (for example, in investigating orbital structures or velocity dispersion anisotropy), and the relationship between the phases comes closer to one-to-one.  The case then for keeping the phases separate appears to reduce.  It  must be remembered that the M2M method can only weight existing, pre-specified orbits / particles.  It does not as yet include a mechanism to modify the orbits dynamically.  Currently the only way to do so is to change the gravitational potential or the initial conditions of the particles by starting a new model.  In the context of the current discussion, this means invoking the N-body phase.

As with all categorisations, situations arise which do not fit completely into any one category.  Within a M2M model, orbiting the particles under self-gravitation perhaps with an additional (dark matter) potential might be appropriate.  For example, in the currrent investigation, if we had had the capability within our M2M implementation, it would have removed the need to construct interpolation tables.  Whether the self-gravitation of particles is regarded as an integral part of M2M modelling or the absorption of N-body modelling into M2M modelling is open to debate.

For the fourth option, conceptually at least, without the benefit of practical experience, adding a M2M capability to an existing N-body software package would appear to be straightforward.  Extrapolating, it should also be straightforward to add a capability to an existing N-body / gas dynamics code, that is to include a M2M style of handing observables.

Regardless of the approach adopted, there is still the need to determine how the two types of modelling should be interfaced.  Certainly initially, this is seen as a manual process requiring assessment of the modelling so far and a judgement as to whether or not the next phase should be invoked.  Automation becomes possible once the `rules of engagement' become clear.  In addition, quite how time is to be regarded must be addressed.  In N-body models, it is not unusual for the evolution of the model with (astrophysical) time to be important, for example in understanding how galactic bars are created and evolve.  In a M2M model, time is no more than part of the mechanism used in adapting the particle weights so that observations of a real galaxy, taken at some specified astrophysical time, are reproduced. Also, dynamical stability over time of the resulting models will need to be considered.

To conclude this section, we have identified some alterative approaches for combining N-body and M2M modelling, and pointed out some of the issues which will need to be addressed in future investigations.

\section{Conclusions}\label{sec:conclusions}
We have met the objectives we set out in the introduction, section \ref{sec:introduction}.  We have extended the made-to-measure method to handle Galactic field observations involving non-parallel projections, and used it with rotating frame kinematics to model the Galactic bar.  Utilising the \citet{Shen2010} N-body particle model of the Galactic bar and bulge, and kinematic data from the BRAVA survey, we have used the made-to-measure method to determine the best fit bar angle and pattern speed matching the data.  Our results are consistent with \citet{Shen2010}.  Whilst this is pleasing, it was not wholly unexpected given the construction of the M2M models. The rotating frame orbit integration is not specific to the Milky Way and could be used with any rotating galaxy.  The \citet{Shen2010} bar pattern speed of $\approx 40 \: \rm{km/s/kpc}$, corroborated by the M2M models in this paper, is lower than the usually quoted value of $60 \: \rm{km/s/kpc}$ for the Milky Way. It is unclear whether such a lower pattern speed is consistent with different observations such as gas dynamics \citep{Bissantz2003} and local kinematics \citep{Dehnen2000OLR, Liu2012}.  The \citet{Shen2010} pattern speed is however close to the value ($42 \: \rm{km/s/kpc}$) determined by \citet{Weiner1999} using gas kinematics, as is the bar angle ($34^{\circ}$).

Regularisation in the M2M method was not used and from our modelling results was not required.  This does not mean that we do not expect to use regularisation in future modelling.  We will decide based on the requirements of the investigation to be performed and the results we are achieving.  Our implementation of the M2M method would benefit from having a particle self-gravitation capability to avoid the need to take a `snap shot' potential from an N-body system, and to avoid the use of interpolation tables.  We will address this in the near future. 

Future simple extensions to the M2M method include modelling the individual stellar velocity measurements, rather than field averages, by enhancing the theory in \citet{Long2010} to handle a dark matter contribution to the line-of-sight velocity distribution.  It would be interesting to understand the impact, if any, on the results of modelling with field averages and individual measurements both in combination and separately.  A related extension would be to include proper motion data, for example from the Optical Gravitational Lensing Experiment (OGLE) \citep{Sumi2004}, in M2M models of the Galaxy.  Such extensions will be valuable once data from the impending GAIA satellite become available. It is not necessary however to wait for GAIA before utilising the extensions to produce improved models of the Milky Way.

We have presented a brief analysis of the options and issues involved in combining N-body and M2M modelling, and have taken a first step in that direction.  Of the issues, the dynamical stability of the resulting model may prove to be the most challenging.  For example, the model of \citet{Wang2012} is able to reproduce the BRAVA kinematic observations but appears to be unstable on timescales longer than 1 Gyr.

\section*{Acknowledgements}
The authors wish to thank the anonymous reviewer for the prompt production of their very helpful comments.  Computer runs were performed on the \textit{Laohu} high performance computer cluster of the National Astronomical Observatories, Chinese Academy of Sciences (NAOC). RJL and SM acknowledge the financial support of the Chinese Academy of Sciences and NAOC. The research presented here is also partially supported by the National Natural Science Foundation of China under grant numbers 11073037 (JS), Y011061001 and Y122071001 (YGW) and by the 973 Program of China under grant number 2009CB824800 (JS).

\bibliographystyle{mn2e}
\bibliography{rjlmodel}

\label{lastpage}
\end{document}